\documentclass{IEEEcsmag}

\usepackage[colorlinks,urlcolor=blue,linkcolor=blue,citecolor=blue]{hyperref}
\expandafter\def\expandafter\UrlBreaks\expandafter{\UrlBreaks\do\/\do\*\do\-\do\~\do\'\do\"\do\-}
\usepackage{upmath,color}
\usepackage{xspace}
\usepackage{subcaption}
\usepackage{booktabs}
\usepackage{graphicx} 
\usepackage[table]{xcolor}

\newcommand{\one}{({\em i}\/)\xspace}
\newcommand{\two}{({\em ii}\/)\xspace}
\newcommand{\three}{({\em iii}\/)\xspace}
\newcommand{\four}{({\em iv}\/)\xspace}

\newcommand{\revOne}[1]{\textcolor{blue}{#1}}

\begin{document}

\sptitle{Regular: Multimodal Interaction and Immersive Multimedia}


\title{Multimodal Cyber-physical Interaction in XR: Hybrid Doctoral Thesis Defense}

\author{Ahmad Alhilal}
\affil{%
   Hong Kong University of Science and Technology, Hong Kong\\
   \phantom{-------------------} Aalto University, Espoo, Finland
}

\author{ Kit Yung Lam}
\affil{%
    \protect Lucerne School of Computer Science and Information Technology, Rotkreuz, Switzerland
}

\author{\protect Lik-Hang Lee}
\affil{%
   Hong Kong Polytechnic University, Hong Kong
}

\author{Xuetong Wang}
\affil{%
    Hong Kong University of Science and Technology, Hong Kong
}

\author{Sijia Li}
\affil{%
 Hong Kong University of Science and Technology (Guangzhou), Guangzhou, China
}

\author{Matti Siekkinen}
\affil{%
   Aalto University, Espoo, Finland
}

\author{Tristan Braud}
\affil{%
   Hong Kong University of Science and Technology, Hong Kong
}

\author{Pan Hui}
\affil{%
Hong Kong University of Science and Technology (Guangzhou), Guangzhou, China\\
\phantom{------------} Hong Kong University of Science and Technology, Hong Kong 
}


\begin{abstract}
Academic events, such as a doctoral thesis defense, are typically limited to either physical co-location or flat video conferencing, resulting in rigid participation formats and fragmented presence. We present a multimodal framework that breaks this binary by supporting a spectrum of participation - from in-person attendance to immersive virtual reality (VR) or browser access - and report our findings from using it to organize the first ever hybrid doctoral thesis defense using extended reality (XR). The framework integrates full-body motion tracking to synchronize the user's avatar motions and gestures, enabling natural interaction with onsite participants as well as body language and gestures with remote attendees in the virtual world. It leverages WebXR to provide cross-platform and instant accessibility with easy setup. User feedback analysis reveals positive VR experiences and demonstrates the framework's effectiveness in supporting various hybrid event activities.

\end{abstract}

\maketitle

\begin{figure*}[!t]
    \centering
  \includegraphics[width=\textwidth]{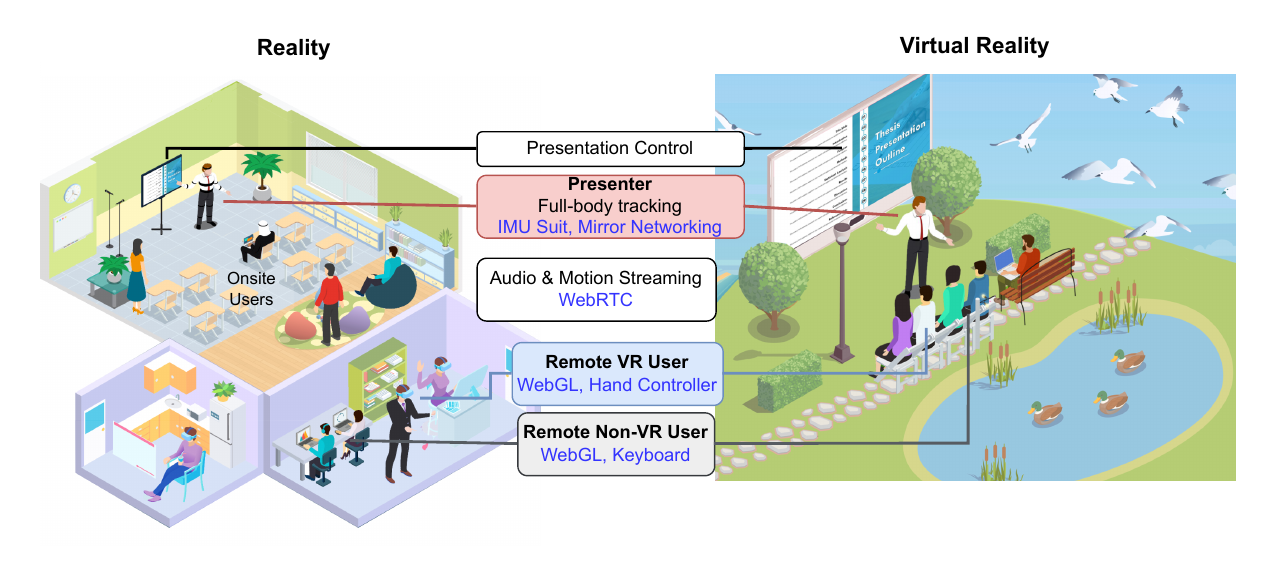}
  \caption{A pictorial description of organizing a hybrid thesis defense with multimodal cyber-physical interaction.}
  \label{fig:teaser}
\end{figure*}
\chapteri{S}ocial events are increasingly transitioning to virtual or hybrid formats. Extended reality (XR) enhances these experiences by creating shared virtual spaces where participants interact through avatars~\cite{metaverseFromMMcomm2023MMmag, towardMetaverse2022MMmag}, combining the flexibility of remote access with the richness of in-person interaction. This capability proves particularly valuable for formal academic events like thesis defenses~\cite{Dermawan2020AnalysisOT}, where maintaining ceremonial gravitas while including remote participants is essential. This article reports on our implementation of the first hybrid doctoral thesis defense, which took place at the Hong Kong University of Science and Technology in 2023.

Traditional thesis defenses have long been confined to physical presence, requiring candidates, examiners, and audiences to gather in a shared location. While traditional thesis defenses foster formality, direct interaction, and nuanced nonverbal communication, they impose geographic and logistical constraints. Moreover, choosing a thesis examination committee (TEC) poses difficulties for candidates, especially when supervisors or external reviewers cannot attend in person due to other obligations. 

The rise of remote collaboration tools has introduced virtual alternatives~\cite{Dermawan2020AnalysisOT}, yet these often fall short in replicating the dynamics of in-person defenses—struggling with rigid participation modes, limited nonverbal expressiveness, and a fragmented sense of presence. With growing pressure for students to complete their degrees on schedule, hybrid defense formats have become essential for addressing geographic limitations while preserving rigorous evaluation timelines. When it comes to academic defenses, seeing is believing. From the viewer's perspective, full-body avatars uniquely preserve the rich nonverbal communication essential for meaningful academic exchanges - from subtle postural shifts to directional cues that upper-body avatars simply cannot convey~\cite{MetaHorizonAvatars}. Thus, realistic virtual presence and communication are crucial to enhance the candidate's message clarity and audience connection. Committee members participate in defenses infrequently, so joining must be effortless through an intuitive, zero-learning interface. The participation should be barrier-free without the need for VR hardware or additional software installations~\cite{netTraffic2023metasys,hybridMetEvent2023kirill}. This is beneficial for time-constrained examiners and supervisors while also encouraging broader attendance among postgraduate students. By addressing these requirements, remote participants can experience the candidate's communication and presentation with similar fidelity as on-site attendees.

In this article, we present our framework (see \autoref{fig:teaser}) for the first hybrid thesis defense, combining in-person attendance with immersive VR participation (via WebXR/WebRTC) and full-body avatar synchronization.
Our framework advances beyond existing video conferencing and social VR platforms through three key contributions: \one \textbf{\textit{Human-avatar interaction framework}}: Full-body motion tracking and replication for natural nonverbal communication, building on our prior work~\cite{hai2022Lam}, \two \textbf{\textit{Hybrid Event Architecture}}: Native support for concurrent physical-virtual participation, preserving the formality and dynamics of in-person defenses while offering geographic flexibility, and \three \textbf{\textit{Barrier-Free Access}}: Intuitive remote participation via WebXR and WebRTC technology requiring no prior technical preparation, eliminating adoption hurdles (e.g., specialized hardware or software expertise).
Thus, the hybrid format preserves the rigor of traditional defenses through structured examination and natural non-verbal cues, while upholding ceremonial formality (e.g., formal avatar attire and full-body gesture) and spatial interactions - ensuring similar scholarly gravitas across physical and virtual participants.

In the rest of the article, we study the virtual alternatives and discuss their limitations (Section~\ref{sec:bkground}). We then present our multimodal interaction framework used in an actual MR hybrid thesis defense (Section~\ref{sec:impl}). Next, we evaluate user perception of the framework's performance and its impact on delivering various activities of thesis defenses (Section~\ref{sec:userstudy}). In Section~\ref{sec:discussion}, we discuss how the framework accommodates the various stages of the hybrid thesis defense.

\begin{table*}[!t]
\centering
\scriptsize
\caption{Comparison of Social VR Platforms by Accessibility, Preparation, and Avatar Features}
\label{tab:vr_comparison}
\begin{tabular}{lllll}
\toprule
\textbf{Platform} & \textbf{Accessibility} & \textbf{Preparation} & \textbf{Avatar Type} \\
\midrule

\textbf{Rec Room} \ref{recroom} & Cross-platform (VR/desktop/mobile) & Account setup, Avatar customization & Full-body, Upper-body,\\
\addlinespace
 & VR headset or desktop client required &  & Cartoony style\\
\midrule
\textbf{VRChat} \ref{vrchat} & VR headset or desktop client required &Account setup, Avatar customization & Full-body (user-generated) \\
\midrule
\textbf{Meta Workrooms} \ref{workrooms} & Meta Quest headset required & Meta account, Workspace setup, & Upper-body,  \\
\addlinespace
 &  & Avatar customization & Precise hand tracking\\
\midrule
\textbf{Spatial} \ref{spatial} & Cross-platform (VR/Desktop/Mobile)  & Account setup, Room config, & Full-body (user-generated) \\
& Browser access & Avatar customization &  \\
\midrule
\rowcolor{blue!10} \textbf{Our Platform} & \textbf{Cross-platform (VR/Desktop/Mobile)} & \textbf{One-click} & \textbf{Full-body (photo-based)} \\
\rowcolor{blue!10}  & \textbf{Browser access} & \revOne{Curated Avatar Creation} &  \textbf{Full-body tracking}\\

\bottomrule
\end{tabular}
\end{table*}

\section{Background and Challenges}
\label{sec:bkground}


We first look at the unique challenges of hybrid thesis defenses, and then discuss existing platforms, their suitability, and limitations.

\subsection{Challenges}
Since TEC members typically participate in defenses occasionally, joining the event must be intuitive using convenient tools, eliminating any learning curve. The technology should facilitate effective verbal and nonverbal communication remotely. Crucially, the technology should remain unobtrusive, allowing the candidate to focus fully on engaging with onsite attendees.

\textbf{Deployment challenges:} Most audience members and TEC members do not possess VR headsets and lack the time to install client apps on their devices, whether VR, mobile, or PC. They expect to join the event without prior preparation, and any technical barriers may discourage them from attending. These additional requirements can hinder their participation and limit their engagement in the defense process. Thesis examination committee members often have demanding schedules and participate
in defenses infrequently. Therefore, joining the virtual thesis defense should be intuitive and require minimal effort.

\textbf{Presenter Engagement:} Thesis defenses are candidate-centered events, where smooth communication and interaction are crucial. Thus, the PhD candidate's speech and physical motions should be accurately captured and replicated in the virtual world. To ensure a high level of engagement, it's essential to accurately capture and recreate the PhD candidate's physical motions. Body language and gestures—such as hand movements and body posture—are powerful tools for conveying ideas and enhancing communication. These gestures add depth and clarity to the message, helping to connect with the audience and making key points more memorable~\cite{gestureForComm2023acmTransGraphx}. Simultaneously, the candidate must maintain full engagement with onsite examiners to preserve the traditional defense's rigor (structured examination), formality (ceremonial protocols), and authenticity (undivided attention). 

\subsection{Background}

\textbf{Video Conferencing Platforms and Hybrid Participation.} Since the COVID-19 pandemic, video conferencing platforms (VCPs) such as Zoom, Microsoft Teams, and Google Meet have become widely adopted for remote academic activities, including thesis defenses. Higher education institutions such as Columbia University have implemented Hybrid-Flexible (HyFlex) learning—a multimodal approach that integrates in-person and online VCP-based participation. This model provides students the flexibility to choose their preferred mode of engagement (onsite, synchronous remote, or asynchronous) for each session while maintaining consistent learning outcomes~\cite{beatty2019hyflex}. Among other institutions, Columbia University Information Technology (CUIT) has deployed integrated audiovisual systems - including video cameras and microphones - across campus classrooms to establish HyFlex learning environments that support concurrent in-person and remote participation~\cite{ColumbiaCTL2023HyFlex}.
While HyFlex VCP-based thesis defenses could facilitate physical-virtual participation, they also impose costs. Candidates must divert valuable time and attention from critical oral presentation preparation and final thesis refinements to manage hybrid logistics. Additionally, administrators face costs related to providing technology-equipped defense rooms and securing extra financial resources~\cite{ColumbiaCTL2023HyFlex}. Moreover, monitoring participant engagement during the thesis defense is difficult due to multiple attendee windows and limited visibility of nonverbal cues from constrained camera angles or screen sharing~\cite{raes2020systematic}.
Moreover, virtual communication norms can disrupt natural interactions, leading to unclear speaking turns and missed opportunities for questions or interjections. Furthermore, remote participants must install and become familiar with the VCP, which often interrupts the flow and formality of the dissertation defenses~\cite{remoteDissertaion2022}.

\textbf{Social VR Platforms.} \autoref{tab:vr_comparison} compares the leading social VR platforms based on accessibility, preparation, and avatar features to determine their suitability for thesis defense environments. Rec Room~\footnote{\url{https://recroom.com/}\label{recroom}} offers game-centric social spaces but lacks essential features—such as a structured turn-taking protocol, formal avatars, and document sharing—necessary for academic defenses. VRChat~\footnote{\url{https://hello.vrchat.com/}\label{vrchat}}, Horizon Workrooms,\footnote{\url{https://www.oculus.com/workrooms}\label{workrooms}}, and Spatial~\footnote{\url{https://spatial.io/}\label{spatial}} can replicate the thesis defense in VR space. Park et al. propose a solution to add richness to non-verbal communications by exploring the use of hand gestures for communicative and control purposes in virtual meetings~\cite{effectiveMeeting2025,JollyGesture2024}.
However, these platforms impose four key participation barriers: (1) hardware and software requirements (VRChat and Rec Room require VR headsets or desktop clients, Horizon Workrooms requires Meta Quest headset),  (2) mandatory account creation, (3) room configuration, and (4) avatar customization by all participants~\cite{netTraffic2023metasys,hybridMetEvent2023kirill}. Our framework reduces these barriers by: \one \textbf{\textit{Centralizing setup:}} Offloading virtual room configuration and avatar generation solely to the presenter; \two \textbf{\textit{Automating identity:}} Using photo-based full-body avatars that require no manual participant customization; and \three \textbf{\textit{Simplifying access:}} Enabling one-click browser joining without pre-installed software.

This article introduces a novel framework that pioneered the first XR thesis defense, combining physical and virtual participation modes.
Our framework specifically addresses time constraints and technical literacy challenges common among users (TEC members and audience), while maintaining full VR capabilities for presenters, such as full-body tracking with real-time avatar coordination.

\section{Real-World Hybrid Thesis Defense Implementation}
\label{sec:impl}
\begin{figure}[!t]
    \centering
    \begin{subfigure}[b] {.5\textwidth}
    \centering
    \includegraphics[width=\linewidth]{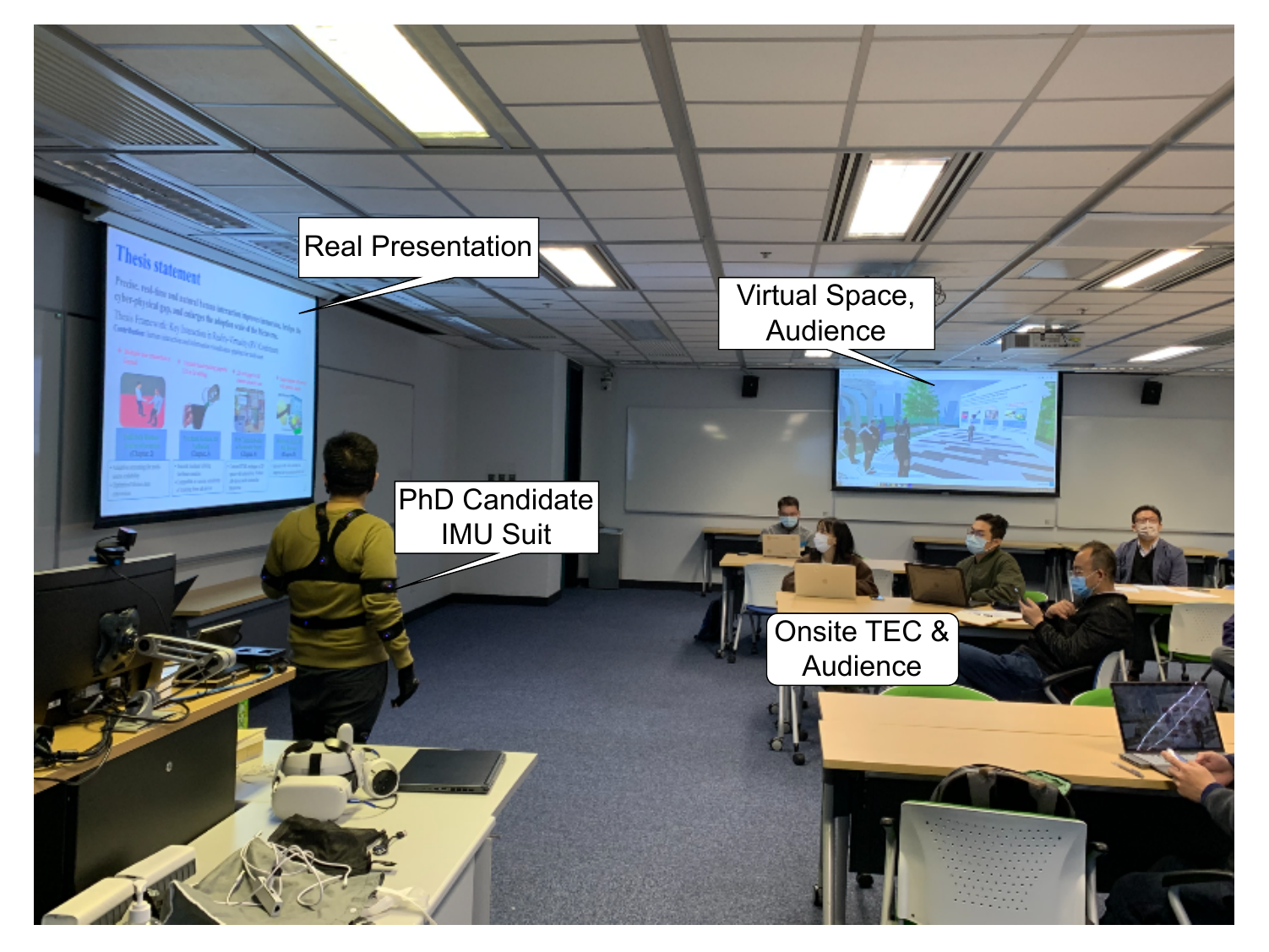}
    \caption{Onsite thesis defense, showing the PhD candidate wearing IMU suit, examiners, audience, one monitor for the real presentation, and another displaying the virtual space.}
    \label{fig:onsiteSpace}
    \end{subfigure}
    \begin{subfigure}[b]{.5\textwidth}
        \centering
\includegraphics[width=\linewidth]{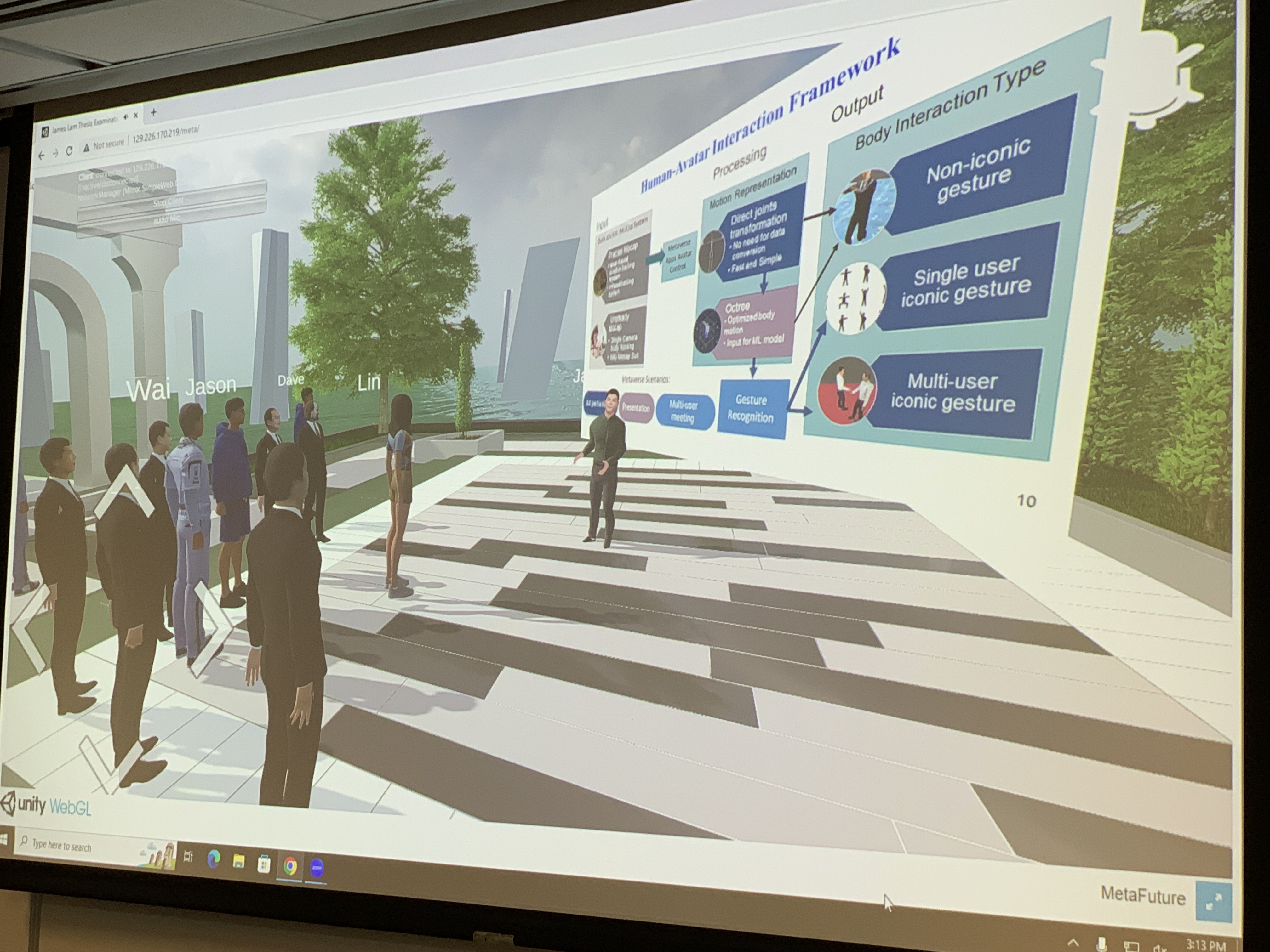}
    \caption{The monitor displayed the virtual space including the presentation, the presenter's avatar, and other users' avatars.}
    \label{fig:VirtualSpace}
    \end{subfigure}

    \caption{Real-world Mixed-reality PhD Defense at Hong Kong University of Science and Technology.}
    \label{fig:MRspace}
\end{figure}

This section presents our implementation of the first Ph.D. defense conducted in a mixed reality environment, integrating three participant roles—presenter, TEC member, and audience—across two modalities (onsite and remote).  As shown in \autoref{fig:teaser} and \autoref{fig:MRspace}, the framework concurrently facilitates both physical participation (onsite attendees in a traditional defense room) and virtual participation (remote attendees via immersive avatar embodiments in a virtual space). 

\subsection{Doctoral Thesis Defense Process}
A typical thesis defense in Hong Kong consists of the candidate's oral presentation, followed with an open discussion with TEC members and the audience, a closed discussion with the TEC members, and deliberations without the candidate. The final decision is communicated by the TEC chair, often concluded with a ceremonial handshake. 
For the oral presentation, the candidate’s IMU suit captures full-body motion, enabling remote attendees to observe natural gestures while maintaining direct eye contact and interaction with onsite examiners.
For open and closed discussions, the spatial audio-visual capabilities in VR enable locating questioning participants, replicating physical examination room dynamics. The handshake between the TEC chair and the candidate is replicated in the virtual world to restore ceremonial formalities. Following, we discuss the technologies and setup to replicate these experiences remotely in VR.

\subsection{Physical Setup and User Roles}
\label{sec:real}
\autoref{fig:teaser} and \autoref{fig:MRspace} illustrate the role-specific devices and hardware configurations (e.g., presenter’s motion capture suit, remote audience interfaces) and the multimodal data streams (audio, motion) captured during the hybrid MR thesis defense.
\autoref{fig:onsiteSpace} depicts the physical setup for the onsite thesis defense, which integrates dual displays with a primary monitor showing the candidate's presentation and a secondary monitor rendering the virtual defense environment. The PhD candidate wears an IMU suit to synchronize full-body avatar motions in the virtual space. This setup enables onsite attendees to view both presentation content (primary monitor) and virtual interactions (secondary monitor) simultaneously, while WebRTC audio streaming (detailed later) facilitates real-time communication with remote participants.
The relay server was hosted on the cloud to relay the users' multimodal data to the remote users.   

We next discuss the four participant roles, categorized by participation mode (onsite/remote) and access type (VR headset/browser-based).  

\textbf{Presenter} is the Ph.D. candidate who defended his thesis. They possess the highest level of interaction in this event. In the defence discussed in this article, we used a Perception Neuron 3 Body Kit tracking device\footnote{\url{https://neuronmocap.com/pages/perception-neuron-3}} with 17 Neuron sensors. The sensors were attached to his head, shoulders, spine, and upper and lower limbs. They tracked and captured the candidate's body motion and gestures. The 17 sensors correspond to 17 sets of position and rotation data, which were processed by a C\# script over Unity3D and mapped to 59 joints of the presenter's avatar to control its body gesture. Additionally, the candidate holds a remote pointer to control the progress of the PowerPoint presentation (PPT) during the defense. \textit{This leads to multimodal data: transform data (position and rotation) for 59 avatar joints, PPT-related commands, and audio}. 

\textbf{Remote VR users} refer to participants who attend the virtual defense remotely using VR headsets. They access the virtual venue (see~\autoref{fig:VirtualSpace}) and use hand controllers to control the motion of their avatars. The application in VR mode applies Inverse Kinematics (IK) to calculate the joint orientations that are calculated backward from selected positions in space. The motions of the avatar's upper limbs are calculated based on the position of the avatar's hands.
Like Meta Workrooms, the application uses movement speed to trigger a lower-body animation- standing or walking- as legs are untracked.
Ultimately, the avatar's gesture and transform are represented in the form of transform data for 9 IK joints. As such, the remote VR users stream two types of data: \textit{transform data for 9 IK joints, lower-body animation state, and audio}.

\textbf{Remote non-VR users} refer to participants who join via web browsers on laptops. They navigate using keyboard inputs (WASD) to move their avatars and mice to adjust their orientation. Because their hands and legs are untracked, the application dynamically blends pre-set animations based on input intensity: idle stance during no input or walking with arm swings for keypresses. These users stream three types of data: \textit{transform data (position and rotation), motion and animation state, and audio}.

\textbf{Onsite participants} attended the defense in person, including the TEC members, supervisors, and general audience. The audio of the onsite participants was recorded and transmitted alongside the presenter's voice.

\subsection{Implementation}
\label{sec:virtual}
We here detail the technical implementation, incorporating role-specific avatars via Avatar SDK, real-time body gesture synchronization via Mirror, audio streaming via WebRTC, and cross-platform via WebXR.

\textbf{Avatar} We designed role-specific avatars for the doctoral thesis defense, generating photorealistic 3D avatars of presenters, supervisors, and examiners that accurately matched their real-world appearance. Prior to each defense session, we collected photographs from the examiners and supervisor through a dedicated web portal, then utilized Avatar SDK~\footnote{\url{https://avatarsdk.com/}} to generate realistic 3D avatars from these photos. Upon login, each examiner and supervisor was automatically assigned their corresponding personalized avatar. For audience members, we provided a curated selection of pre-designed, generic human-like avatars from which they could choose upon entering the virtual environment. Once any user—whether examiner, supervisor, or audience—logged into the system, their avatar was automatically distributed to all other connected clients, enabling all participants to see and interact with each other's visual representations in the shared virtual space. The doctoral defense being candidate-focused, the presenter's avatar was pre-loaded into all participants' local XR environments by default.


\textbf{Data Transmission/Synchronization} The framework implements avatar and slide synchronization using Mirror\footnote{\url{https://mirror-networking.com/}}, a free and open-source game networking library. It streams the presenter's full-body motion (59 joint transforms) and active slide, VR users' IK data (9 joints) and lower-body animation state, and non-VR users' transforms and limb animation state.  Remote applications process this multimodal data to render animated avatars in real-time.

\textbf{Virtual Slideshow} We converted PowerPoint slides into individual images and imported them as textures in Unity3D. We then created a material using these textures and mapped it onto a VR plane, representing the presentation display. Finally, implemented a slideshow C\# script to navigate through the textures using keyboard inputs for 'Next' and 'Previous' controls. The presenter's VR application transmitted the slide number to synchronize across all connected clients in real-time. The presenter controlled the virtual and real slides simultaneously for consistency between the two realities.

\textbf{Audio} The framework utilizes WebRTC\footnote{\url{https://webrtc.org}} for audio communication with remote users. It enables a bidirectional audio channel to allow Q\&A during the open Q\&A session. In the defense discussed in this article, all participants—including examiners, supervisors, and remote users (both VR and non-VR)—were muted by default. Onsite attendees (examiners, supervisors, and audience) shared the candidate’s physical space, so their audio was captured and transmitted along with the presenter’s voice. During the Q\&A session, the users could unmute and ask questions.

\textbf{Cross-platform and Web Access} The framework leverages WebXR API~\footnote{\url{https://immersiveweb.dev/}} to deliver VR experiences on web browsers. This provides native cross-platform XR support, eliminating the need for platform-specific applications (desktop or VR). For motion data processing in our Unity3D environment (C\# code), it integrates the SimpleWebXR library~\footnote{\url{https://github.com/Rufus31415/Simple-WebXR-Unity}} to interface between Unity3D and the WebXR JavaScript API. Remote participants using VR headsets can join directly through a WebXR-compatible browser, with hand controllers enabling natural interactions like virtual hand movement. SimpleWebXR library uses WebSockets for communication between WebXR and Unity3D, while WebGL delivers hardware-accelerated rendering, resulting in a fully functional VR experience accessible through standard web browsers.

\section{User Study}
\label{sec:userstudy}

\begin{figure*}[!t]
    \centering
    \begin{subfigure}[b]{.39\textwidth}
    \centering
    \includegraphics[width=\linewidth]{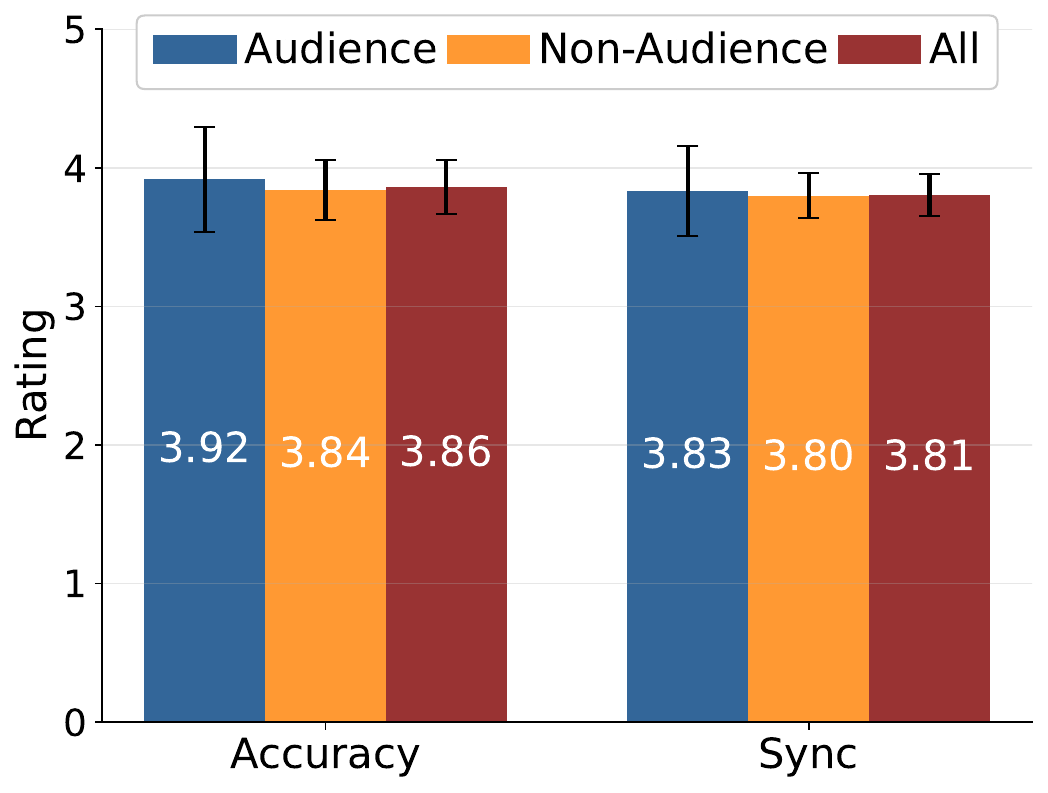}
        \caption{ Perception of performance.}
        \label{fig:perfQoE_grouped}
    \end{subfigure}
    \begin{subfigure}[b]{.6\textwidth}
        \centering
        \includegraphics[width=\linewidth]{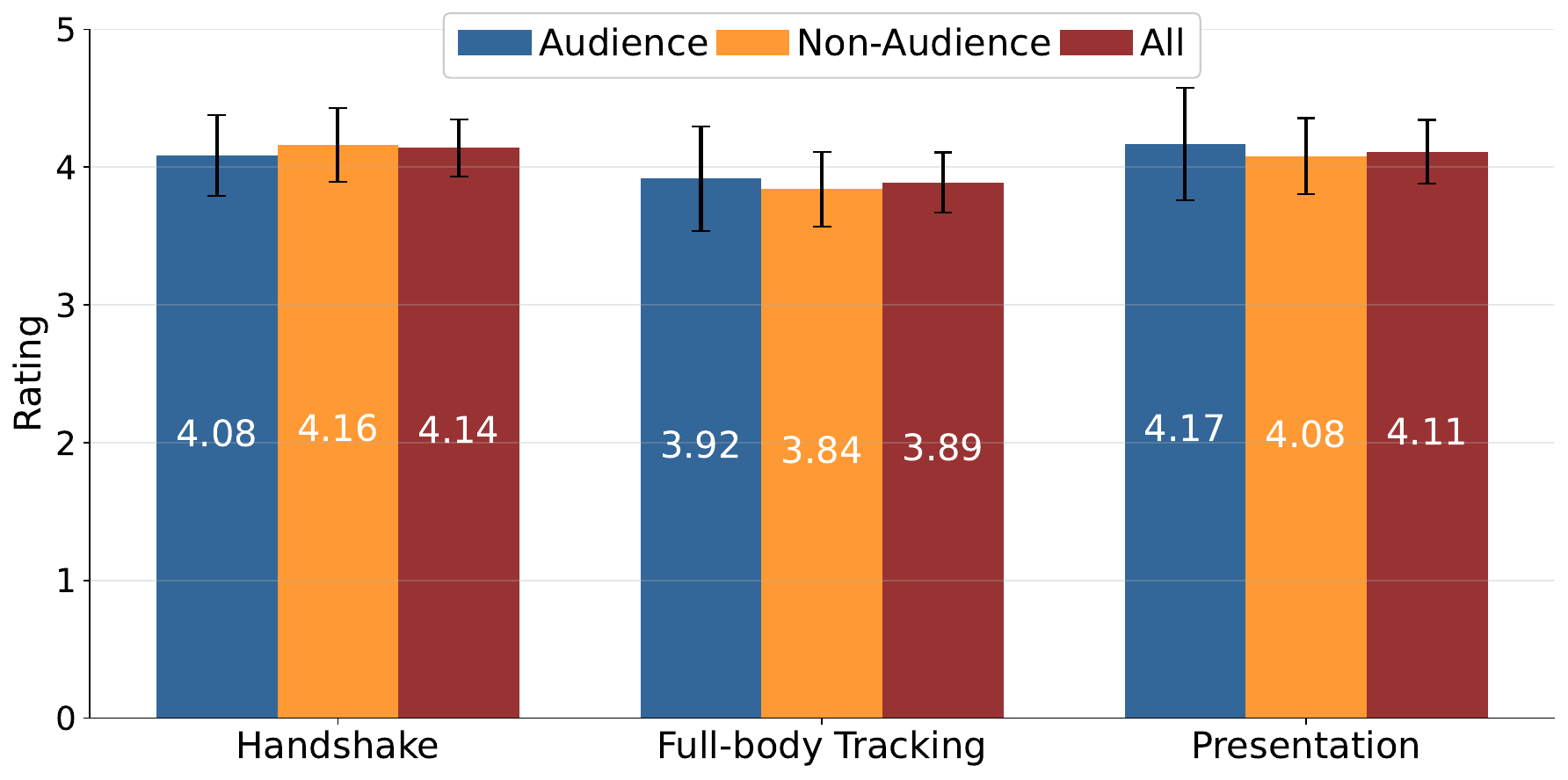}
        \caption{ Perception of functionality for defense activities.}
        \label{fig:promisQoE_grouped}
    \end{subfigure}
    \caption{User-perceived MR thesis defense performance: attendee ratings (n=12, in-person/remote) vs. non-attendee ratings (n=24).}
    \label{fig:qoe}
\end{figure*}

We conducted a user study to evaluate participant perceptions of the virtual Ph.D. defense, with particular emphasis on how candidates' performances were perceived in the digital environment.

\subsection{\textbf{Participants. }}
We administered an online survey to 36 participants (12 defense attendees, 24 non-attendees), comprising 22 males and 14 females aged 19-42 years. They are classified according to their familiarity with the technology into the following categories: unfamiliar (2/36), slightly familiar (9/36), moderately familiar (12/36), and very familiar (13/36). The survey aimed to collect their feedback on the framework.

\subsection{\textbf{Procedure. }}
As the doctoral defense centers on the candidate, our evaluation focused on participants' perceptions of the candidate's virtual performance, including formal multi-user interactions like handshaking. During the concluding announcement stage, an onsite examiner wore a VR headset, shook hands with and congratulated the candidate, blending physical and virtual elements. Attendees (n=12) then rated their experience relative to an in-person defense. 

Non-attendees (n=24) watched a video of a real person wearing the IMU suit along with the synchronized avatar motion for three representative use cases in the VR, \one \textbf{Handshake}: two people shaking hands, \two \textbf{Presentation}: A person giving a presentation using PowerPoint slides, and \three \textbf{Full-body Tracking}: A person moving in a freestyle (hand gesture and limbs motion). After watching each video, using an online survey, the participants filled out a Likert scale survey on their experience. We asked all the study participants to rate the perceived accuracy of the avatar recreation on a 5-point scale, ranging from very inaccurate to very accurate: \textit{(1: very inaccurate, 2: inaccurate, 3: somewhat accurate, 4: accurate, 5: very accurate)}.
They rate the synchronization between the physical body motion and the avatar's motion on a 5-point scale, ranging from no synchronization to fully synchronized: \textit{(1: no synchronization at all, 2: no synchronization, 3: somewhat synchronized, 4: synchronized, and 5: fully synchronized)}.
Since the participants have experience in XR technologies, they also rated the promise of the multimodal interaction to enable a successful virtual defense over each use case on a 5-point scale, ranging from 'definitely not' to 'definitely yes'. 

\subsection{\textbf{Results. }}

Figure~\ref{fig:qoe} presents the comparative results across participant groups (attendees, non-attendees, and all participants), showing mean ratings with 95\% confidence intervals. The analysis shows consistently positive experiences among all participants, with particularly high ratings from the Audience group (attendees): accuracy (3.92/5), synchronization (3.83/5), handshake (4.08/5), full-body tracking (3.92/5), and presentation (4.17/5). Non-Audience group (non-attendees) reported similarly favorable outcomes: accuracy (3.84/5), sync (3.80/5), handshake (4.16/5), full-body tracking (3.84/5), and presentation (4.08/5), while the aggregate results across all participants remained robust: accuracy (3.86/5), sync (3.81/5), handshake (4.14/5), full-body tracking (3.89/5), and presentation (4.11/5).

Overall, the results indicate a favorable perception of the performance in the VR space. Participants reported a high perceived accuracy (M=3.86, 95\% CI $\pm$0.15), and high synchronization between the presenter's and avatar's motions (M=3.81, 95\% CI $\pm$0.16). The MR thesis defense platform achieves robust motion accuracy and synchronization, confirming reliable real-time tracking and avatar-presenter coordination. These metrics indicate that the platform successfully minimizes drift, which is crucial for natural nonverbal communication akin to in-person interactions, thereby enhancing engagement in VR.

The Activity-specific evaluations further highlight the MR thesis defense’s effectiveness and suitability for essential defense activities: \one Handshake interactions (M=4.14, 95\% CI ±0.14) scored highest, successfully replicating natural social gestures -- particularly during ceremonial moments (e.g., thesis committee chair congratulating candidates); \two Full-body tracking (M=3.89, 95\% CI ±0.17) performed consistently well, supporting dynamic movements without perceptible lag for natural nonverbal communication akin to in-person interactions; and \three Presentation functionality (M=4.11, 95\% CI ±0.13) received the most stable ratings, highlighting the platform's reliability for core tasks that support academic discourse during virtual defenses, including slide navigation, annotation, and figure/line referencing.

These findings collectively validate our multimodal framework's efficacy in supporting all critical aspects of virtual thesis defenses, demonstrating both technical robustness (all metrics > 3.8/5) and operational reliability (95\% CIs $\leq \pm$0.17). In terms of universal accessibility reflected in participant feedback,  multiple attendees noted they could join the thesis defense \textit{"immediately after clicking the link—no tutorials or setup needed."} One committee member remarked, \textit{"I expected technical hurdles, but it worked seamlessly through my Chrome browser"}, underscoring the elimination of specialized hardware requirements. These experiences align with our design goal of enabling truly barrier-free participation, where even first-time users could engage without preparatory steps.

\section{Discussion}
\label{sec:discussion}

\subsection{Process Summary}

We start by analyzing how our framework streamlines the doctoral thesis defense process and captures participants' feedback on its effectiveness.

\textbf{Oral Presentation.} The high-fidelity motion capture enables natural body language expression during presentations, improving communication clarity and audience engagement in the virtual environment. Importantly, the IMU-based approach eliminates the need for VR headsets or desktop interfaces, allowing the candidate to maintain direct eye contact and interaction with onsite examiners.

\textbf{Open/Closed Discussion.} The VR environment enables presenters to instantly locate questioning participants, replicating physical examination room dynamics through: (1) spatial 3D audio that facilitates locating the speaker in 3D space, and (2) an optimized visual layout allowing complete visibility of all avatars through natural head movements.

\textbf{Outcome Announcement.} Upon successful completion of the defense, the committee chair may perform a virtual or in-person handshake with the candidate. User study participants responded favorably to this gesture and full-body tracking in VR. To engage remote audiences, the framework transmits real-time body and hand movement data of both the committee chair and the candidate over the internet to remote users' applications. The applications then render synchronized handshake animations between the avatars. This preserves the ceremonial formality of the VR defense—a critical advantage over conventional video platforms that lack spatial interactions.

\subsection{Performance and Perception} 
Our framework integrates full-body avatar interaction with a presenter-facing monitor. This enables natural non-verbal communication with remote participants. Our user study revealed a favorable reception of the framework's performance, with participants noting the high accuracy and synchronization of the presenter's avatar motions. Activity-specific evaluations confirmed the framework's effectiveness for core defense tasks, with high ratings for virtual presentations, full-body tracking accuracy, and ceremonial handshaking. The candidate did not need a VR headset or desktop interface, enabling them to focus fully on the physical defense space while their motions and gestures were replicated and rendered in VR. This preserved the candidate’s undivided attention for onsite attendees while simultaneously engaging remote attendees. 

\subsection{Advancing SoTA.} The framework bridges the gaps of video conferencing platforms – including low perceived formality, inadequate nonverbal communication, and poor attendee visibility. In particular, it incorporates four key solutions: \one the immersive 3D environment recreates physical defense settings through photorealistic 3D avatars-based interactions, restoring ceremonial formality; \two embodied interactions enabled by full-body gesture tracking and spatial audio preserve nonverbal communication cues like demonstrative gestures for illustrating concepts, emphatic motions for highlighting key points, and intentional postural shifts; \three the 3D spatial and high-resolution audio replicate real-world acoustic fidelity; and \four natural 3D interactions: Questioning participants move closer with voice/avatar size adjusting by proximity—maintaining full visibility while eliminating rigid video grids. 
Our framework reduces setup complexity compared to HyFlex~\cite{ColumbiaCTL2023HyFlex}, requiring only a dual-display configuration—one screen for the real presentation and another for the virtual environment. Unlike HyFlex, which relies on multiple cameras, microphones, and specialized audiovisual infrastructure, our approach minimizes hardware dependencies.
Our framework also advances social VR platforms by enabling presenter-avatar interaction, concurrent physical-virtual participation, and intuitive barrier-free access via zero-install web technologies.

\subsection{Limitations and Future Directions }
The presenter-avatar framework does not capture facial expressions, diminishing emotional connection with remote interactions compared to in-person interactions.
The virtual slideshow implementation has two key limitations. First, the image-based conversion of PPT slides loses interactive elements (animations, transitions, and embedded media). Second, requiring simultaneous control of physical and virtual slides increases the presenter's cognitive load and risks a mismatch between the two realities.
The framework requires presenters to wear and carefully calibrate an IMU suit, adding technical complexity that may disrupt the defense's natural progression. While the suit itself isn't uncomfortable, this setup process demands time and technical attention that could otherwise be focused on presentation delivery.
Vision-based approaches like 3D Gaussian splatting (TaoAvatar~\cite{Chen_2025_CVPR}) could eliminate wearable sensors, enabling automatic full-body capture through cameras alone. This would make the process completely unnoticeable to onsite participants.

\section{Conclusion}
\label{sec:conclusion}
This article presents our multimodal XR framework for hybrid doctoral thesis defenses, the first to enable remote participation. The framework incorporates WebXR technology for cross-platform accessibility with no technical prerequisites, enabling universal participation. The framework combines real-time motion tracking for full-body avatar synchronization with WebRTC-powered audio streaming, enabling natural verbal and non-verbal communication. This dual-channel approach replicates core defense activities - from oral presentations and natural gestures to formal ceremonies - within VR without compromising the authenticity of in-person interactions. Our user study revealed positive perceptions of the framework's performance and its ability to maintain engagement across physical and virtual spaces. Future improvements will focus on replacing IMU suits with vision-based capture, supporting interactive virtual slides, and adding facial tracking to strengthen emotional connection—streamlining the framework while enhancing realism.

\section{Acknowledgments}
This research was partially supported by a grant from the Guangzhou Municipal Nansha District Science and Technology Bureau under Contract No. 2022ZD01 and the MetaHKUST project from the Hong Kong University of Science and Technology (Guangzhou). It was also partially supported by a grant from the Hong Kong Innovation and Technology Commission under the Innovation and Technology Fund, Grant Number ITS/319/22FP. We gratefully acknowledge AvatarSDK for their generous sponsorship, which provided one-year access to their avatar generation API.

\def\refname{References}

\bibliographystyle{IEEEtran}
\bibliography{references} 

\begin{IEEEbiography}{Ahmad Alhilal}{\,} is a postdoc researcher at Aalto University, Finland. He was a postdoc fellow at the Hong Kong University of Science and Technology (HKUST). His current research interests include Network-aware streaming, Scalable Computing, and Cloud-rendered Mobile/XR Gaming. Dr. Alhilal received his Ph.D. degree in Computer Science and Engineering from HKUST. He is a member of the IEEE Computer Society. Contact him at ahmad.alhilal@aalto.fi.\vspace*{8pt}
\end{IEEEbiography}

\begin{IEEEbiography}{Kit Yung Lam}{\,} is a research associate at Lucerne School of Computer Science and Information Technology, Rotkreuz, Switzerland. His research interests include augmented reality, multimodal interaction, and immersive media experiences. Dr. Lam received his PhD in Computer Science and Engineering from Hong Kong University of Science and Technology. Contact him at kityung.lam@hslu.ch.
\end{IEEEbiography}

\begin{IEEEbiography}{Lik-Hang Lee}{\,} is an assistant professor at Hong Kong Polytechnic University, Hong Kong SAR. His current research interests include mixed reality, human-computer interaction, LLM, and education technologies. Dr. Lee received his Ph.D. degree in Computer Science and Engineering from Hong Kong University of Science and Technology. He is an IEEE senior member. Contact him at lik-hang.lee@polyu.edu.hk\vspace*{8pt}
\end{IEEEbiography}

\begin{IEEEbiography}{Xuetong Wang}{\,} is a Ph.D. candidate at The Hong Kong University of Science and Technology (HKUST). Her research interests include human-computer interaction, extended reality, and artificial intelligence. Xuetong received her Bachelor of Engineering in Electronic Engineering from HKUST. Contact her at xwangdd@connect.ust.hk.\vspace*{8pt}
\end{IEEEbiography}

\begin{IEEEbiography}{Sijia Li}{\,} is an independent researcher in Cambridge, UK. Her research interests include HCI, AI for Education, and Efficient Learning. Sijia received her Master's degree in Computational Media and Arts from Hong Kong University of Science and Technology (Guangzhou). Contact her at sli308@connect.hkust-gz.edu.cn.\vspace*{8pt}
\end{IEEEbiography}

\begin{IEEEbiography}{Matti Siekkinen}{\,} is a university lecturer at Aalto University, Espoo, Finland. His research interests include cloud-accelerated Virtual and Augmented Reality, distributed multimedia systems, and mobile computing. Dr. Siekkinen received his Ph.D. degree in Computer Science from Eurecom / University of Nice Sophia Antipolis, France. Contact him at matti.siekkinen@aalto.fi.\vspace*{8pt}
\end{IEEEbiography}

\begin{IEEEbiography}{Tristan Braud}{\,} is an Assistant Professor in the Division of Integrative Systems and Design at the Hong Kong University of Science and Technology (HKUST), where he leads the Extended Reality and Immersive Media (XRIM) Lab. His research explores the intersection of ubiquitous computing and human-centered systems design, with a focus on augmented and virtual reality as the primary interaction medium. Dr. Braud received his Ph.D. degree in Computer Science from Université Grenoble Alpes, France. Contact him at braudt@ust.hk\vspace*{8pt}
\end{IEEEbiography}

\begin{IEEEbiography}{Pan Hui}{\,} is a Chair Professor of Computational Media and Arts and Director of the Centre for Metaverse and Computational Creativity at the Hong Kong University of Science and Technology (Guangzhou), and a Chair Professor of Emerging Interdisciplinary Areas at the Hong Kong University of Science and Technology. His research interests include  Data-driven Systems Design and Immersive Human-Data Interaction. Prof. Hui received his PhD from the Computer Laboratory at the University of Cambridge. He is an International Fellow of the Royal Academy of Engineering (FREng), a Member of the Academia Europaea (MAE), an IEEE Fellow, and a Distinguished Scientist of the Association for Computing Machinery (ACM). Contact him at panhui@ust.hk.\vspace*{8pt}
\end{IEEEbiography}

\end{document}